\documentclass[letterpaper,journal]{IEEEtran}
\usepackage{amsmath,amsfonts}
\usepackage{algorithmic}
\usepackage{algorithm}
\usepackage{array}
\usepackage{textcomp}
\usepackage{stfloats}
\usepackage{url}
\usepackage{verbatim}
\usepackage{graphicx}

\usepackage{xcolor}

\usepackage[german,ngerman,english]{babel}%
\PassOptionsToPackage{%
    babel,%
    style=english%
}{csquotes}%
\addto{\captionsenglish}{%
  \renewcommand*{\contentsname}{Content}%
  \renewcommand*{\listfigurename}{Figures}%
  \renewcommand*{\listtablename}{Tables}%
  \renewcommand*{\figurename}{Figure}%
  \renewcommand*{\tablename}{Table}%
}%

\PassOptionsToPackage{%
	colorlinks=false,%
	urlcolor=black,%
	linkcolor=black,%
	citecolor=black,%
	filecolor=black,
	breaklinks,%
}{hyperref}%
\usepackage[%
	pdfencoding=auto,%
	pdfborder={0 0 0},%
]{hyperref}%

\usepackage[utf8]{inputenc}%
\usepackage[T1]{fontenc}

\usepackage{csquotes}%

\usepackage[%
	acronym=true,
	nonumberlist,%
	nopostdot,%
	seeautonumberlist,%
	shortcuts,%
	section=chapter,%
	toc=false,%
	nogroupskip=true,%
    prefix,
]{glossaries-extra}%
\setabbreviationstyle[acronym]{long-short}%
\glsdisablehyper
\newacronym{nn}{NN}{Nomadic Network}
\newacronym{leo}{LEO}{Low Earth Orbit}
\newacronym{geo}{GEO}{Geosynchronous Earth Orbit}
\newacronym{isl}{ISL}{inter-satellite link}
\newacronym{ue}{UE}{user equipment}

\newacronym{ar}{AR}{Application Relation}%
\newacronym{indeco}{InDeCo}{Interconnection Detachment \& Convergence}
\newacronym{coma}{CoMa}{Context-Management}
\newacronym{peac}{PeAc}{Permission \& Access}
\newacronym{dynploy}{DynPloy}{Dynamic Deployment}
\newacronym{remcon}{RemCon}{Remote Configuration}
\newacronym{coagd}{CAD}{Collective Aggregator Discovery}
\newacronym{archdaemon}{ArchDaemon}{Architecture-Daemon}
\newacronym{ctrllog}{ACL}{Aggregated Control Logic}
\newacronym{adagent}{ADAg}{ArchDaemon-Agent}
\newacronym{adaloc}{ADAl}{ArchDaemon-Allocator}
\newacronym{inserv}{InServ}{Inserted-Service}
\newacronym{infrserv}{InfrServ}{Infrastructure-Service}
\newacronym{atserv}{AtServ}{Atop-Service}%
\newacronym{orcus}{OrCuS}{Organic Customization System}
\newacronym{ore}{ORE}{Organic Runtime Environment}%
\newacronym{relag}{Agent}{Agent}%
\newacronym{relorc}{Orc}{Orchestrator}%
\newacronym{slaag}{SLAAg}{Service Level Agreement Agent}%
\newacronym{cmdc}{CMDC}{Centralized Mission and Driving Control}%
\newacronym{dcmod}{DCMod}{Data Collection Module}%
\newacronym{posmod}{PosMod}{Positioning Module}%
\newacronym{mts}{MTS}{Map Transformation Stage}%
\newacronym{sensmod}{SensMod}{Sensing Module}%
\newacronym{sidr}{SIDR}{Sensor ID Resolver}%
\newacronym{camod}{CAMod}{Carrier Aggregation Module}%
\newacronym{muxs}{MuxS}{Multiplexer Stage}%
\newacronym[plural=CSEs,%
	firstplural=Carrier State Estimators (CSEs),%
	longplural=Carrier State Estimators]%
	{cse}{CSE}{Carrier State Estimator}%
\newacronym{ca}{CA}{Carrier Aggregator}%
\newacronym[plural=IAs,%
	firstplural=Information Anchors (IAs),%
	longplural=Information Anchors]%
	{ia}{IA}{Information Anchor}%
\newacronym{dbmod}{DBMod}{Data Base Module}%
\newacronym{bui}{BUI}{Base User Interface}%
\newacronym{bman}{BMan}{Base Manager}%
\newacronym[plural=DSs,%
	firstplural=Data Suppliers (DSs),%
	longplural=Data Suppliers]%
	{ds}{DS}{Data Supplier}%
\newacronym[plural=GWs,%
	firstplural=Gateways (GWs),%
	longplural=Gateways]%
	{gw}{GW}{Gateway}%
\newacronym{auditf}{TTF}{Trust \& Traceability Function}%
\newacronym{specal}{SASF}{Spectrum Allocation \& Sharing Function}%
\newacronym{gcl}{GCL}{Gate Control List}
\newacronym{gnb}{gNB}{gNodeB}
\newacronym{mac}{MAC}{Medium Access Control}
\newacronym{qos}{QoS}{Quality-of-Service}%
\newacronym{utc}{UTC}{Coordinated Universal Time}%
\newacronym[plural=KPIs,%
	firstplural=Key Performance Indicators (KPIs),%
	longplural=Key Performance Indicators]%
	{kpi}{KPI}{Key Performance Indicator}%
\newacronym{poc}{PoC}{Proof-of-Concept}%
\newacronym{wip}{WiP}{Work in Progress}%
\newacronym{utf}{UTF}{Unicode Transformation Format}%
\newacronym{inft}{IT}{Information Technology}%
\newacronym{fsm}{FSM}{Finite-State Machine}%
\newacronym{fsa}{FSA}{Finite-State Automaton}%
\newacronym{fst}{FST}{Finite-State Transducer}
\newacronym{umlsm}{UMLSM}{UML State Machine}%
\newacronym{uml}{UML}{Unified Modeling Language}%
\newacronym[longplural={Distributed Ledger Technologies}]{dlt}{DLT}{Distributed Ledger Technology}%
\newacronym{gps}{GPS}{Global Positioning System}%
\newacronym{trl}{TRL}{Technology Readiness Level}%
\newacronym{opsys}{OS}{Operating System}%
\newacronym{digt}{DT}{Digital Twin}%
\newacronym{rand}{R\&D}{Research \& Development}%
\newacronym{softdr}{SDR}{Software Defined Radio}%
\newacronym{sota}{SotA}{State-of-the-Art}%
\newacronym{api}{API}{Application Programming Interface}%
\newacronym{uri}{URI}{Uniform Resource Identifier}%
\newacronym{netfunc}{NF}{Network Function}%
\newacronym{nfv}{NFV}{Network Function Virtualization}%
\newacronym{typlv}{TLV}{Type-Length-Value}%
\newacronym{loc}{LoC}{Lines of Code}%
\newacronym{url}{URL}{Uniform Resource Locator}%
\newacronym{nic}{NIC}{Network Interface Card}%
\newacronym{rest}{REST}{Representational State Transfer}%
\newacronym{ftp}{FTP}{File Transfer Protocol}%
\newacronym{http}{HTTP}{Hypertext Transfer Protocol}%
\newacronym{https}{HTTPS}{Hypertext Transfer Protocol Secure}%
\newacronym{rpc}{RPC}{Remote Procedure Call}%
\newacronym{lldp}{LLDP}{Link Layer Discovery Protocol}%
\newacronym{snmp}{SNMP}{Simple Network Management Protocol}%
\newacronym{nsh}{NSH}{Network Service Header}%
\newacronym{sdn}{SDN}{Software-defined Networking}%
\newacronym[plural=SDNs,%
	firstplural=Software-defined Networks (SDNs),%
	longplural=Software-defined Networks]%
	{sdnet}{SDN}{Software-defined Network}%
\newacronym{ip}{IP}{Internet Protocol}%
\glsunset{ip}%
\newacronym{tcp}{TCP}{Transmission Control Protocol}%
\newacronym{udp}{UDP}{User Datagram Protocol}%
\newacronym{dns}{DNS}{Domain Name System}%
\newacronym{ntp}{NTP}{Network Time Protocol}%
\newacronym{ptp}{PTP}{Precision Time Protocol}%
\newacronym{p2p}{P2P}{Point-to-Point}%
\newacronym{peer2p}{Peer2P}{Peer-to-Peer}%
\newacronym{m2m}{M2M}{Machine-to-Machine}%
\newacronym{rbs}{RBS}{Reference Broadcast Time Synchronization}%
\newacronym{rbis}{RBIS}{Reference Broadcast Infrastructure Synchronization}%
\newacronym{mdns}{mDNS}{Multicast DNS}%
\newacronym{dnssd}{DNS-SD}{DNS Service Discovery}%
\newacronym{servoa}{SOA}{Service-Oriented Architecture}%
\newacronym{rtt}{RTT}{Round Trip Time}%
\newacronym{csp}{CSP}{Communication Service Provider}%
\newacronym{mmtc}{mMTC}{Massive Machine Type Communication}%
\newacronym{arq}{ARQ}{Automatic Repeat Request}%
\newacronym{ppm}{PPM}{Parallel Process Migration}%
\newacronym{checkres}{C/R}{Checkpoint/Restore}%
\newacronym{plmn}{PLMN}{Public Local Mobile Network}%
\newacronym{snr}{SNR}{Signal to Noise Ratio}%
\newacronym{acp}{AP}{Access-Point}%
\newacronym{urllc}{uRLLC}{Ultra Reliable Low Latency Communication}%
\newacronym{bnetza}{BNetzA}{Bundesnetzagentur}%
\newacronym{mno}{MNO}{Mobile Network Operator}
\newacronym{msp}{MSP}{Mobile Service Provider}
\newacronym{5g}{5G}{fifth-generation}%
\newacronym{6g}{6G}{sixth-generation}%
\newacronym{bey5g}{B5G}{Beyond 5G}%
\newacronym{5gsba}{SBA}{5G Service Based Architecture}%
\newacronym{sba}{SBA}{Service Based Architecture}%
\newacronym{mano}{MANO}{Management \& Orchestration}%
\newacronym{ran}{RAN}{Radio Access Network}%
\newacronym{cn}{CN}{Core Network}
\newacronym{cran}{C-RAN}{Centralized - Radio Access Network}%
\newacronym{oran}{O-RAN}{Open - Radio Access Network}%
\newacronym{sbi}{SBI}{Service Based Interface}%
\newacronym{jcas}{JC\&S}{Joint Communication and Sensing}%
\newacronym{rrm}{RRM}{Radio Resource Management}%
\newacronym{uequ}{UE}{User Equipment}%
\newacronym{gmlc}{GMLC}{Gateway Mobile Location Center}%
\newacronym{embb}{eMBB}{enhanced Mobile Broad-band}%
\newacronym{mnoper}{MNO}{Mobile Network Operator}%
\newacronym{coretc}{CtC}{Core-to-Core}%
\newacronym{gnodb}{gNB}{5G gNodeB Basestation}%
\newacronym{3gpp}{3GPP}{3rd Generation Partnership Project}%
\newacronym{uwb}{UWB}{Ultra Wideband}%
\newacronym{newr}{NR}{New Radio}%
\newacronym{runit}{RU}{Radio Unit}%
\newacronym{cunit}{CU}{Centralized Unit}%
\newacronym{dunit}{DU}{Distributed Unit}%
\newacronym{cpl}{CP}{Control Plane}%
\newacronym{upl}{UP}{User Plane}%
\newacronym{sdr}{SDR}{Software Defined Radio}
\newacronym{nsaas}{NSaaS}{Network-Slicing-as-a-Service}
\newacronym{naas}{NaaS}{Network-as-a-Service}
\newacronym{taas}{TaaS}{Trust-as-a-Service}
\newacronym{haps}{HAPS}{High Altitude Platform Station}
\newacronym{npn}{NPN}{non-public network}
\newacronym{nnpn}{NNPN}{nomadic non-public network}
\newacronym{pc-nnpn}{PC-NNPN}{predictable, cross-border nomadic non-public network}
\newacronym{ss-nnpn}{SS-NNPN}{scheduled, single operator nomadic non-public network}
\newacronym{uus-nnpn}{UUS-NNPN}{unpredictable, unscheduled, and safety-relevant nomadic non-public network}

\newacronym{cbrs}{CBRS}{Citizens Broadband Radio Service}
\newacronym{ntn}{NTN}{Non-terrestrial network}
\newacronym{ngap}{NGAP}{Next Generation Application Protocol}%
\newacronym{amf}{AMF}{Access and Mobility Management Function}%
\newacronym{upf}{UPF}{User Plane Function}%
\newacronym{lmf}{LMF}{Localization Management Function}%
\newacronym{nef}{NEF}{Network Exposure Function}%
\newacronym{nrf}{NRF}{Network Repository Function}%
\newacronym{scp}{SCP}{Service Communication Proxy}%
\newacronym{manet}{MANet}{Mobile Ad hoc Network}
\newacronym{mvno}{MVNO}{Mobile Virtual Network Operator}
\newacronym{sdo}{SDO}{Standards Developing Organization}

\newacronym{i40}{I4.0}{Industry 4.0}%
\newacronym{cps}{CPS}{Cyber-Physical System}%
\newacronym{plc}{PLC}{Programmable Logic Controller}%
\newacronym{vpc}{VPC}{Virtual Process Controller}%
\newacronym{tsn}{TSN}{Time-Sensitive Networking}%
\newacronym{cnc}{CNC}{Central Network Controller}%
\newacronym{cuc}{CUC}{Central User Configuration}%
\newacronym{tsncnc}{TSN CNC}{TSN Central Network Controller}%
\newacronym{tssdn}{TSSDN}{Time-Sensitive Software-Defined Networking}%
\newacronym{cpci}{CPCI}{Central Process Control Instance}%
\newacronym{optech}{OT}{Operation Technology}%
\newacronym{hmtc}{HMTC}{Human-Machine Type-Communication}
\newacronym{iot}{IoT}{Internet of Things}%
\newacronym{iiot}{IIoT}{Industrial Internet of Things}%
\newacronym{agv}{AGV}{Automated Guided Vehicle}%
\newacronym{amr}{AMR}{Autonomous Mobile Robot}%
\newacronym{uav}{UAV}{Unmanned Aerial Vehicle}%
\newacronym{hap}{HAP}{High Altitude Platform}%
\newacronym{erp}{ERP}{Enterprise Resource Planning}%
\newacronym{ips}{IPS}{Indoor Positioning System}%
\newacronym{vmc}{VMC}{Vehicle Motion Control}%
\newacronym{imu}{IMU}{Inertial Measurement Unit}%
\newacronym{ros}{ROS}{Robot Operating System}%
\newacronym{opc}{OPC}{Open Platform Communication}%
\newacronym{opcua}{\gls{opc} UA}{OPC Unified Architecture}%
\newacronym{mqtt}{MQTT}{Message Queue Telemetry Transport}%
\newacronym{etl}{ETL}{Extract, Transfer, Load}%
\newacronym{dds}{DDS}{Data Distribution Service}%
\newacronym{dgps}{DGPS}{Differential Global Positioning System}%
\newacronym{ecef}{ECEF}{Earth-centered, Earth-fixed}%
\newacronym{gnss}{GNSS}{Global Navigation Satellite System}%
\newacronym{arti}{AI}{Artificial Intelligence}%
\newacronym{mlearn}{ML}{Machine Learning}%
\newacronym{svm}{SVM}{Support Vector Machine}%
\newacronym{sla}{SLA}{Service Level Agreement}%
\newacronym{b2b}{B2B}{Business-To-Business}%
\newacronym{b2c}{B2C}{Business-To-Consumer}%
\newacronym{pmse}{PMSE}{Programme Making and Special Events}%
\newacronym{ppdr}{PPDR}{Public Protection and Disaster Relief}%
\usepackage{support_caption}%
\usepackage{caption}%
\usepackage{subcaption}
\usepackage[%
	backend=biber,
	hyperref=true,%
	sorting=none,%
	natbib=true,%
	language=english,%
	defernumbers=true,%
	locallabelwidth,%
	pagetracker=false,
	citecounter=false,
	citetracker=true,
]{biblatex}%
\DeclareFieldFormat{sentencecase}{#1}%
\makeatletter%
\let\cbx@citehook=\empty%
\DeclareCiteCommand{\citeff}[\cbx@DKwrapper]
{%
    \usebibmacro{cite:init}%
    \usebibmacro{prenote}%
}%
{%
    \usebibmacro{citeindex}%
    \usebibmacro{cite:comp}%
    \usebibmacro{citeAddFootOnFirst}%
}%
{}
{%
    \usebibmacro{cite:dump}%
    \usebibmacro{postnote}%
}%
\newbibmacro*{citeAddFootOnFirst}{%
    \ifciteseen{}%
    {%
        \xappto\cbx@citehook{%
        \noexpand\footnotetext[\thefield{labelnumber}]{%
            $^{*}$%
            \citefield{\thefield{entrykey}}{title}%
            \ifnameundef{author}{}{ | \citename{\thefield{entrykey}}{author}}%
            \iffieldundef{year}{}{ | \citefield{\thefield{entrykey}}{year}}%
        }}%
    }%
}%
\newrobustcmd{\cbx@superscript}[1]{%
    \mkbibsuperscript{#1}%
    \cbx@citehook%
    \global\let\cbx@citehook=\empty%
}%
\newrobustcmd{\cbx@DKwrapper}[1]{%
    \bibopenbracket%
    #1%
    \cbx@citehook%
    \bibclosebracket%
    \ifx\cbx@citehook\empty%
    \else%
        $^{*}$%
    \fi%
    \global\let\cbx@citehook=\empty%
}%
\makeatother%
\newlength\bibhangDenKr%
\defbibenvironment{bibliographyDenKrEnv}%
{\list%
    {%
        \printtext[labelnumberwidth]{%
        \printfield{labelprefix}%
        \printfield{labelnumber}}%
    }%
    {%
        \setlength{\labelwidth}{\labelnumberwidth}%
        \setlength{\leftmargin}{\labelwidth}%
        \addtolength{\leftmargin}{\labelsep}%
        \setlength{\itemsep}{\bibitemsep}%
        \setlength{\parsep}{\bibparsep}%
    }%
}%
{\endlist}%
{\item}%
\newcommand{\labpragAsym}{\raisebox{.5\height}{\textbf{\scriptsize$\rightarrow$}}}%
\newcommand{\labpragBsym}{\raisebox{.5\height}{\textbf{$\scriptstyle\Rightarrow$}}}%
\newcommand{\labpragCsym}{\raisebox{.5\height}{\scalebox{0.7}{$\pointright$}}}%
\newcommand{\labpragDsym}{\raisebox{.4\height}{\scalebox{0.7}{$\blacktriangleright$}}}%
\newcommand{\labpragEsym}{\raisebox{.1\height}{\scalebox{0.6}{\compoundarrow}}}%
\newcommand{\labDKsymBullet}{\textbullet}%
\newcommand{\labDKsymGeviert}{\textbf{--}}%
\newcommand{\labDKsymAst}{$\mathrm{\ast}$}%
\newcommand{\labDKsymDot}{\textperiodcentered}%
\newcommand{\labDKsymOblong}{\raiseC{\oblongdash}}%
\newcommand{\labDKsymPointright}{\raiseC{\scalebox{0.7}{$\pointright$}}}%
\newcommand{\labDKsymTriangle}{\raisebox{0.3\height}{\scalebox{0.9}{$\triangleright$}}}%
\newcommand{\labDKsymTriangleBl}{\raiseC{\scalebox{0.7}{$\blacktriangleright$}}}%
\newcommand{\labDKsymRTriCurvedB}{\raiseC{\scalebox{1}{$\trianglerightcurvedback$}}}%
\newcommand{\labDKsymLozenge}{\raiseC{\scalebox{0.65}{$\mathrm{\lozenge}$}}}%
\newcommand{\labDKsymLozengeBl}{\raiseC{\scalebox{0.65}{$\mathrm{\blacklozenge}$}}}%
\newcommand{\labDKsymSquare}{\raiseC{\scalebox{0.55}{$\mathrm{\square}$}}}%
\newcommand{\labDKsymSquareBl}{\raiseC{\scalebox{0.55}{$\mathrm{\blacksquare}$}}}%
\newcommand{\labDKsymDiamondSplit}{\raiseC{\scalebox{0.7}{\ding{118}}}}%
\newcommand{\labDKsymCrossA}{\raiseC{\scalebox{0.7}{\ding{103}}}}%
\newcommand{\labDKsymCrossB}{\raiseC{\scalebox{0.7}{\ding{104}}}}%
\newcommand{\labDKsymCrossC}{\raiseC{\scalebox{0.7}{\ding{105}}}}%
\newcommand{\labDKsymCrossD}{\raiseC{\scalebox{0.7}{\ding{106}}}}%
\newcommand{\labDKsymCrossE}{\raiseC{\scalebox{0.7}{\ding{107}}}}%

\usepackage[inline]{enumitem}%
\newlist{enuminlrom}{enumerate*}{1}%
\setlist[enuminlrom]{label=\textit{(\roman*)}}%
\newlist{enumparrom}{enumerate}{1}%
\setlist[enumparrom]{%
	label=\textit{\textbf{(}\Roman*\textbf{)}},%
	align=left,%
	labelindent=0.5\parindent,%
	listparindent=\parindent,%
	leftmargin=0pt,%
	labelwidth=!,%
	itemindent=!,%
	topsep=0ex,%
	itemsep=0ex,%
	parsep=0ex,%
}%
\SetEnumitemKey{nolabel}{label=,leftmargin=0pt}%
\setlist[description]{%
	topsep=0ex,%
	itemsep=0ex,%
	parsep=0ex,%
	labelindent=0pt,%
	leftmargin=0pt%
}%
\setlist[itemize]{%
	topsep=0ex,%
	itemsep=0ex,%
	parsep=0ex,%
	leftmargin=*,
}%
\setlist[enumerate]{%
	topsep=0ex,%
	itemsep=0ex,%
	parsep=0ex,%
	leftmargin=1.5em%
}%
\providecommand{\DenKrDescriptionlabelFont}{}%
\providecommand{\DenKrDescriptionlabelFormat}{}%
\providecommand{\DenKrDescriptionlabelMake}{}%
\renewcommand{\DenKrDescriptionlabelFont}{\normalfont\bfseries}%
\renewcommand{\DenKrDescriptionlabelFormat}[1]{#1:}%
\renewcommand{\DenKrDescriptionlabelMake}[1]{\DenKrDescriptionlabelFormat{{\DenKrDescriptionlabelFont#1}}}%
\renewcommand{\descriptionlabel}[1]{\hspace{\labelsep}\DenKrDescriptionlabelMake{#1}}%
\SetEnumitemKey{labItDefaultA}{label=\textbullet,align=right,leftmargin=*}%
\SetEnumitemKey{labItDefaultB}{label=\textbf{--},align=right,leftmargin=*}%
\SetEnumitemKey{labItDefaultC}{label=$\mathrm{\ast}$,align=right,leftmargin=*}%
\SetEnumitemKey{labItDefaultD}{label=\textperiodcentered,align=right,leftmargin=*}%
\SetEnumitemKey{labpragA}{label=\labpragAsym,align=right,leftmargin=1.3em}
\SetEnumitemKey{labpragB}{label=\labpragBsym,align=right,leftmargin=1.3em}
\SetEnumitemKey{labpragC}{label=\labpragCsym,align=right,leftmargin=1.3em}
\SetEnumitemKey{labpragD}{label=\labpragDsym,align=right,leftmargin=1.3em}
\SetEnumitemKey{labpragE}{label=\labpragEsym,align=right,leftmargin=1.3em}
\SetEnumitemKey{labDKBullet}{label=\labDKsymBullet,align=right,leftmargin=*}
\SetEnumitemKey{labDKGeviert}{label=\labDKsymGeviert,align=right,leftmargin=*}
\SetEnumitemKey{labDKAst}{label=\labDKsymAst,align=right,leftmargin=*}
\SetEnumitemKey{labDKDot}{label=\labDKsymDot,align=right,leftmargin=*}
\SetEnumitemKey{labDKOblong}{label=\labDKsymOblong,align=right,leftmargin=*}%
\SetEnumitemKey{labDKPointright}{label=\labDKsymPointright,align=right,leftmargin=*}
\SetEnumitemKey{labDKTriangle}{label=\labDKsymTriangle,align=right,leftmargin=*}
\SetEnumitemKey{labDKTriangleBl}{label=\labDKsymTriangleBl,align=right,leftmargin=*}
\SetEnumitemKey{labDKRTriCurvedB}{label=\labDKsymRTriCurvedB,align=right,leftmargin=*}%
\SetEnumitemKey{labDKLozenge}{label=\labDKsymLozenge,align=right,leftmargin=*}
\SetEnumitemKey{labDKLozengeBl}{label=\labDKsymLozengeBl,align=right,leftmargin=*}
\SetEnumitemKey{labDKSquare}{label=\labDKsymSquare,align=right,leftmargin=*}
\SetEnumitemKey{labDKSquareBl}{label=\labDKsymSquareBl,align=right,leftmargin=*}
\SetEnumitemKey{labDKDiamondSplit}{label=\labDKsymDiamondSplit,align=right,leftmargin=*}
\SetEnumitemKey{labDKCrossA}{label=\labDKsymCrossA,align=right,leftmargin=*}
\SetEnumitemKey{labDKCrossB}{label=\labDKsymCrossB,align=right,leftmargin=*}
\SetEnumitemKey{labDKCrossC}{label=\labDKsymCrossC,align=right,leftmargin=*}
\SetEnumitemKey{labDKCrossD}{label=\labDKsymCrossD,align=right,leftmargin=*}
\SetEnumitemKey{labDKCrossE}{label=\labDKsymCrossE,align=right,leftmargin=*}
\newlist{itemizeDefault}{itemize}{4}%
\newlist{enumerateDefault}{enumerate}{4}%
\setlistdepth{9}%
\renewlist{itemize}{itemize}{9}%
\setlist[itemize,1]{labDKBullet,leftmargin=*}%
\setlist[itemize,2]{labDKOblong,leftmargin=*}%
\setlist[itemize,3]{labDKLozengeBl,leftmargin=*}%
\setlist[itemize,4]{labDKRTriCurvedB,leftmargin=*}%
\setlist[itemize,5]{labDKDiamondSplit,leftmargin=*}%
\setlist[itemize,6]{labDKAst,leftmargin=0.9em}%
\setlist[itemize,7]{labDKSquareBl,leftmargin=*}%
\setlist[itemize,8]{labDKPointright,leftmargin=1.1em}%
\setlist[itemize,9]{labDKDot,leftmargin=*}%
\renewlist{enumerate}{enumerate}{9}%
\setlist[enumerate,1]{label=\arabic*.}%
\setlist[enumerate,2]{label=\alph*)}%
\setlist[enumerate,3]{label=\roman*.}%
\setlist[enumerate,4]{label=\Alph*)}%
\setlist[enumerate,5]{label=\Roman*.}%
\setlist[enumerate,6]{label=(\arabic*)}%
\setlist[enumerate,7]{label=(\roman*)}%
\setlist[enumerate,8]{label=(\Roman*)}%
\setlist[enumerate,9]{label=(\alph*)}%

\usepackage{fontawesome5}%
\definecolor{orcidlogocol}{HTML}{A6CE39}%
\newcommand{\DenKrAuthorOrcidLogo}[1]{\href{https://orcid.org/#1}{\raisebox{0.5ex}[0pt][0pt]{\resizebox{!}{1.5ex}{\textcolor{orcidlogocol}{\faIcon{orcid}}}}}}%

\begin{document}

\title{Use Cases, Metrics, and Challenges of Nomadic Non-Public Networks for the 6G Standardization}

\author{%
Daniel Lindenschmitt\IEEEauthorrefmark{1}\DenKrAuthorOrcidLogo{0000-0002-4088-7158},
Michael Gundall\IEEEauthorrefmark{2}\DenKrAuthorOrcidLogo{0000-0002-0180-3767},
Ainur Daurembekova\IEEEauthorrefmark{1}\DenKrAuthorOrcidLogo{0009-0004-3620-6120},
Marcos Rates Crippa\IEEEauthorrefmark{1}\DenKrAuthorOrcidLogo{0000-0002-3332-4672},\\
Mohammad Asif Habibi\IEEEauthorrefmark{1}\DenKrAuthorOrcidLogo{0000-0001-9874-0047},
Bin Han\IEEEauthorrefmark{1}\DenKrAuthorOrcidLogo{0000-0003-2086-2487},
Philipp Rosemann\IEEEauthorrefmark{1}\DenKrAuthorOrcidLogo{0009-0007-3135-4601},
Dennis Krummacker\IEEEauthorrefmark{2}\DenKrAuthorOrcidLogo{0000-0001-9799-4870},\\
Benedikt Veith\IEEEauthorrefmark{2}\DenKrAuthorOrcidLogo{0000-0002-5344-1964},
and
Hans D. Schotten\IEEEauthorrefmark{1}\IEEEauthorrefmark{2}\DenKrAuthorOrcidLogo{0000-0001-5005-3635}
\\[\baselineskip]%
\IEEEauthorrefmark{1}RPTU University Kaiserslautern-Landau, Germany.\\%
\IEEEauthorrefmark{2}German Research Center for Artificial Intelligence (DFKI), Germany.\\%
%
%
}


\pagenumbering{gobble} 
\maketitle

\begin{abstract}
\input{"2_abstract".tex}
\end{abstract}

%
%
%
%


\section{Introductory Remarks} \label{sec:intro}

\IEEEPARstart{6}{G} introduces novel use cases~\cite{giordani2020toward}, driven by disruptive functionalities that challenge traditional mobile radio communication architectures~\cite{tataria20216g}. One of these features is the introduction of \glspl{npn} in \gls{5g}, which has already created a fundamental basis for the expansion of technologies beyond traditional mobile network operators~\cite{3gpp-ts23501,3gpp-ts38300}. 
\glspl{npn} allow the owners of these networks to operate independent \gls{5g} networks within specific frequency ranges and limited areas.

On one hand, \glspl{npn} enable the development of innovative applications that leverage mobile communication standards for greater efficiency or introduce new deployment scenarios. On the other hand, their growing adoption is driving increased hardware availability, creating a pull effect that reinforces the significance of \glspl{npn}. Furthermore, \gls{5g} standardization has introduced nomadic nodes~\cite{osseiran2014scenarios}, which expand possible areas of application for \glspl{npn} by enabling flexible network deployment. These single base stations temporarily enhance the network capacity or network coverage~\cite{osseiran2014scenarios}. To address limitations of individual nomadic node solutions, the concept of \glspl{nn} was introduced in ~\cite{6GNomadNet_ArchChal} and recommendations for adaptation with regard to \gls{6g} standardization were elaborated. Since \glspl{nn} are typically operated outside traditional \glspl{mno}, they generally fall under the category of \glspl{npn}, leading to the emergence of \glspl{nnpn}~\cite{lindenschmitt2024nomadic}. 
In the field of ad-hoc or temporary networks, solutions already exist. Examples of these are self-organizing \glspl{manet} or the neutral host concept, which enables third-party providers to manage and share network infrastructure among multiple \glspl{mno}. Another example is the \gls{mvno} approach, where operators do not invest in owning spectrum licenses or \gls{ran} infrastructure but lease network access from \glspl{mno}. In addition \gls{cbrs} is a common concept in the United States, where the Spectrum Access System manages frequency allocation to protect incumbents and priority access licenses~\cite{7158261}. However, none of these concepts provide a 3GPP-compliant solution for \glspl{nn} in which an \gls{npn} is fully functional while on the move. This addressed new approaches in the area of network interfaces and data backhauling, enabling the implementation of novel use cases, e.g., in the area of \gls{ppdr} or the public or private transport sector.

The Recommendation  ITU-R~M.2160~\cite{ITU-R_M.2160-0} plays a key role in standardizing \glspl{nn}, particularly for ubiquitous connectivity. It also introduces new metrics related to coverage and sustainability, essential for establishing \glspl{nnpn}~\cite{10298069}. Among the key enablers for \glspl{nnpn} are \glspl{ntn}, which extend connectivity beyond terrestrial infrastructures~\cite{osseiran2014scenarios}.  While these enablers provide a strong foundation for the development of \gls{nnpn}, several challenges, that require extensive research, remain on the path to fully realized \glspl{nnpn}. One critical challenge is the need for a streamlined and flexible regulatory approval process, which current regulations do not yet accommodate.  

The paper at hand defines \gls{nnpn}-related use cases and distinguishes them from \glspl{npn}. Section~\ref{sec:scenarios} presents use cases that are enabled or significantly improved by \glspl{nnpn}. Section~\ref{sec:metric} introduces \glspl{kpi} and relevant metrics for evaluating these use cases and discusses \gls{nnpn} clustering. Section~\ref{sec:challenges} outlines key challenges in \gls{nnpn} realization. Finally, section~\ref{sec:concl} ends the article and outlines future research directions.

\section{Scenarios of nomadic mobile communication networks}
\label{sec:scenarios}
The introduction of \gls{npn} offers private users and companies the possibility to set up and fine-tune their own wireless networks based on \gls{5g} within particular frequencies. 
With \glspl{npn}, the number of private operators of \gls{5g} networks has increased steadily, and the number of hardware providers and system integrators has also increased significantly.
After the early adoption phase of this transformative technology, companies are increasingly relying on \ac{npn} as the underlying technology for their communication requirements. This results in new requirements and use cases that go beyond the capabilities of the original \ac{npn} idea and are not covered by standardization. This phenomenon is particularly evident in connection with \glspl{nnpn}, which represent a separate form of \ac{npn}. \glspl{nnpn} are characterized by the fact that they overcome the previously mandatory location-based licensing and thus have the freedom to be relocated during operation. 
This poses new challenges for the network interfaces standardized by \glspl{sdo} like \gls{3gpp}, in particular the N2 and N3 interfaces. This is necessary because new functionalities are required to ensure the secure operation of \glspl{nnpn}. An overview of the requirements is summarized in Table~\ref{tab:interface}. The necessary features relate, e.g., to an adapted, dual-mode \gls{amf}, a so-called proxy \gls{upf}, and secure authentication via trust anchors. This extension of the network interfaces enables new types of use cases that were not possible due to the previous limitations of \glspl{npn}. Table~\ref{tab:comp_npn_nnpn} provides an overview of the new capabilities introduced by \glspl{nnpn} and the use cases that benefit from them. Here, the key feature relates to the location-based nature of previous \glspl{npn} and the introduction of the nomadic aspect with \glspl{nnpn} (see location-bound while operating in Table~\ref{tab:comp_npn_nnpn}). However, there are also other aspects, such as the possibility of choosing a hybrid infrastructure approach, in which \gls{ran} and \gls{cn} are provided partly by the \gls{nnpn} itself and partly by \glspl{mno}. This leads to necessary adjustments to the current \gls{3gpp} standardization and regulatory conditions, as summarized in Table~\ref{tab:interface} and Table~\ref{tab:comp_npn_nnpn}. The following subsection discusses the respective requirements of the individual use cases in detail.

\begin{table}[!ht]
\centering
\caption{Necessary adjustments of network interfaces according to \gls{3gpp}}
\label{tab:interface}
\begin{tabular}{|c|c|c|}
\hline
\textbf{Requirement} & \textbf{Interface} & \textbf{Necessary features} \\
\hline
Operating locally &  & Dual-mode N2 with \\
during separation & N2 & embedded \gls{amf}\\
\hline
Coordination of &  & Discovery \& secure   \\
stakeholders & N2 & signaling\\
\hline
User-plan & & Proxy-\gls{upf} and\\
forwarding & N3 & session handling\\
\hline
Dynamic & N2 \& & Transport-aware  \\
connectivity & N3 & signaling\\
\hline
Security \& & N2 \& & Local authentication\\
trust & N3 & with trust anchors \\
\hline
\end{tabular}
\end{table}

       \begin{table*}[!htb]
            \centering
            \caption{Comparison of core functionalities of \glspl{npn} and \glspl{nnpn} and requirements of defined use cases}
            \label{tab:comp_npn_nnpn}
            \begin{tabular}{|c|c|c||c|c|c|c|c|}
             \hline
                          &     &       &       &       &       & \textbf{Construction} & \textbf{Public/private}\\
                          & \textbf{NPN} & \textbf{NNPN} & \textbf{PMSE} & \textbf{PPDR} & \textbf{Agriculture} & \textbf{ sites} & \textbf{transport}\\
            \hline
             \textbf{Limited coverage area} & \checkmark & \checkmark & \checkmark & \checkmark & \checkmark & \checkmark & \checkmark\\
            \hline
             \textbf{Location-bound while operating} & \checkmark & $\times$ & $\sim$ & $\times$ & $\sim$ & $\sim$ & $\times$\\
            \hline
             \textbf{Network slicing option} & \checkmark & $\times$ & $\sim$ & $\times$ & $\sim$ & $\sim$ & $\times$\\
            \hline
             \textbf{Hybrid infrastructure option} &  &  &  &  &  &  &\\
             \textbf{(\gls{ran}-\gls{cn} allocation)} & $\sim$ & \checkmark & \checkmark & \checkmark & \checkmark & \checkmark & \checkmark\\
            \hline
            \textbf{Multiple stakeholders possible} & $\times$ & \checkmark & \checkmark & \checkmark & $\sim$ & $\sim$ & \checkmark\\
            \hline
            \textbf{Requirements covered by \gls{3gpp}} & \checkmark & $\times$ & $\sim$ & $\times$ & $\sim$ & $\sim$ & $\times$\\
            \hline
            \textbf{Regulatory requirements covered} & \checkmark & $\times$ & $\sim$ & $\times$ & $\sim$ & $\sim$ & $\times$\\
            \hline
            \end{tabular}
      \end{table*}

\begin{figure*}[!ht]
    \centering
    \footnotesize
    \subfloat[Visualization of a \gls{nnpn} in a Programme Making and Special Events scenario.]{
        \includegraphics[width=0.7\textwidth]{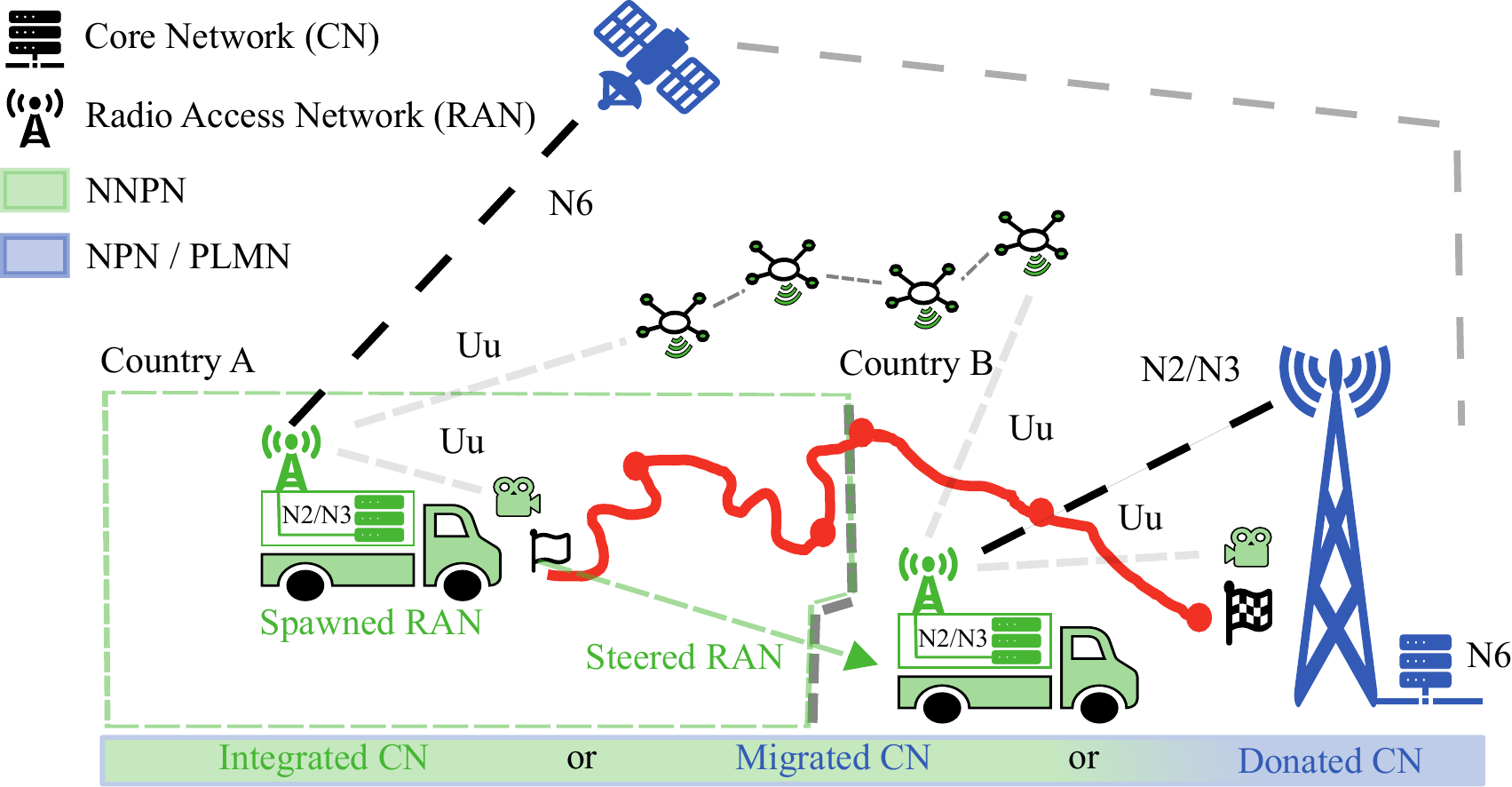}
        \label{fig:pmse}}

    \subfloat[Visualization of a \gls{nnpn} in a Public Protection and Disaster Relief scenario.]{
        \includegraphics[width=0.7\textwidth]{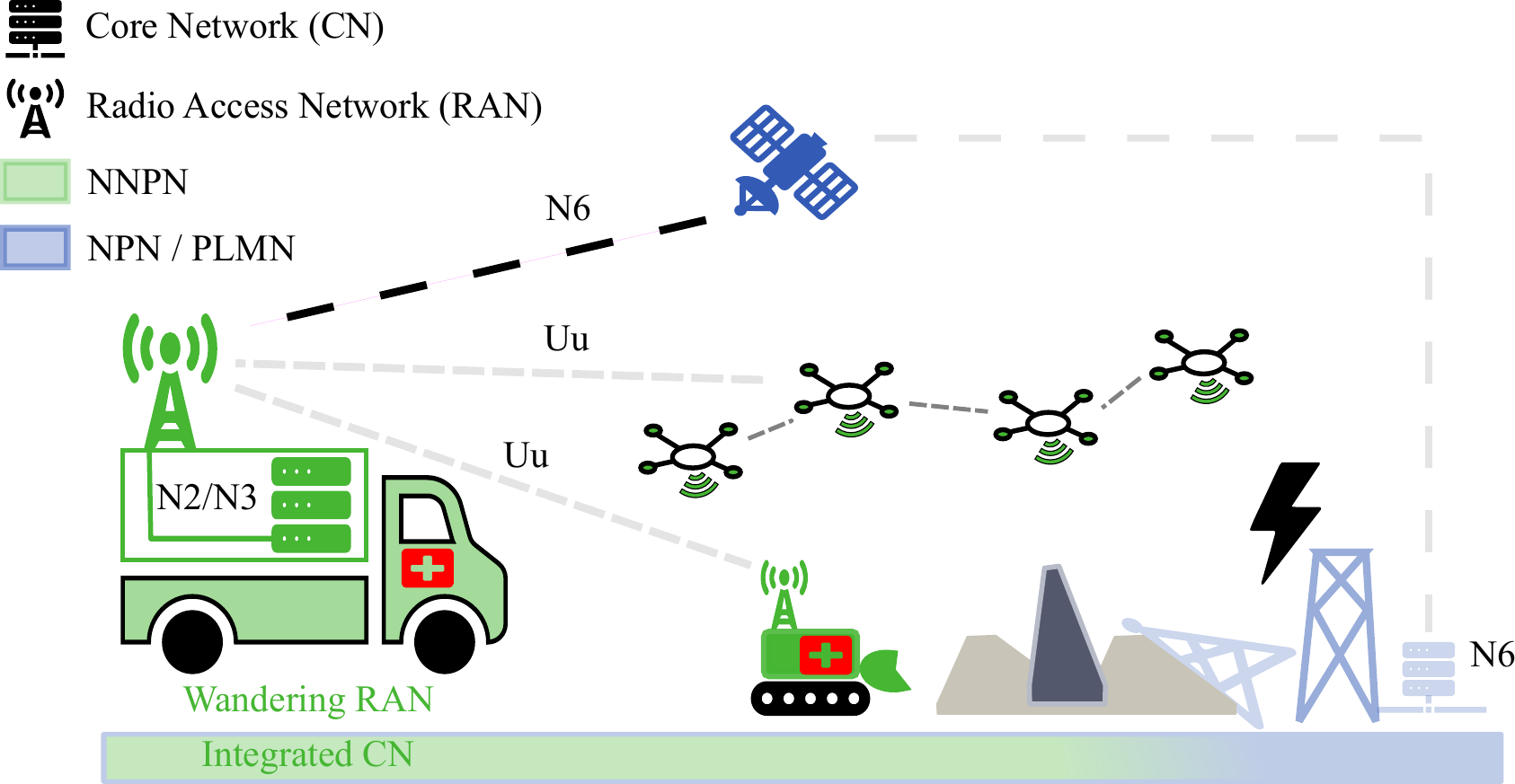}
        \label{fig:ppdr}}
     \caption{Realization scenario of \glspl{nnpn} for Programme Making and Special Events and Public Protection and Disaster Relief}
    \label{fig:pmse_ppdr}
\end{figure*}

\subsection{Use cases in the scope of \glsfmtshortpl{nnpn}}
\label{subsec:usecases}
Evolving to \textit{nomadic} \glspl{npn} enables entirely new applications as it introduces spatial mobility also to the infrastructure components and thereby new capabilities to provide connectivity or support flexible scenarios.

\subsubsection{\gls{pmse}}
A variety of different event categories and news productions in the \gls{pmse} sector make extensive use of radio communication. Events such as elections, sports competitions and similar activities are not limited to single or fixed venues, but take place at different locations and times. In media production, major events are pre-planned, but there are also unforeseen requirements, particularly in the context of news production. News production must react quickly and flexibly to ongoing events and at the same time satisfy the public's need for high-quality reporting that takes place in unprepared and changing locations.
The communication demands and mobility on-site require a wireless communication solution that is under the control of the user, while offering predictable \gls{qos}. This can be well served by \gls{5g} \glspl{npn} for \textit{planned} events.
The spatial and temporal flexibility of \textit{unscheduled} events however, require the communication infrastructure in addition to be transportable. \glspl{nnpn} concepts facilitate this by adopting spectrum licensing procedures and possible cross-border operation \cite{5GrecordsEBU}.

Figure~\ref{fig:pmse} illustrates a \gls{pmse} cross-border scenario. In this example, an \gls{nnpn} is used in Country A, which realizes the necessary connectivity of \glspl{ue} in the coverage area. 
The \gls{ran} is first established as a spawned \gls{ran} at a fixed location, afterwards becoming a steered \gls{ran} which moves along a defined route to Country B. On \gls{cn} side, depending on the use case and environment, different solutions can be implemented. With an \textit{Integrated \gls{cn}}, the entire hardware is located on a vehicle and processed; in addition, data could be transmitted to a central cloud via an \gls{ntn} backhaul using the N6 interface. Furthermore, implementation as a migrated \gls{cn} or donated \gls{cn} as defined in \cite{6GNomadNet_ArchChal} is possible, in which parts or the entire \glspl{netfunc} are outsourced to existing infrastructure of \glspl{npn} or \glspl{plmn} using the N2/N3 interface. 
The extension to this corresponding communication leads to necessary changes to the previous standardization of the interfaces, in addition requirements for licensing and regulation in cross-border activities need to be addressed by the ongoing \gls{6g} standardization as further discussed in Section \ref{sec:challenges}.

\subsubsection{\gls{ppdr}}

This field refers to facilities such as amublance, fire brigades, or emergency services. They face special and often complex challenges every time they are deployed. \gls{5g} provides connectivity for data transmission for \glspl{uav}, ground-based \glspl{agv} and mobile clients. At crowded events such as soccer games, music events or political demonstrations, security officers equipped with network-connected wearables can use remotely controlled, camera-equipped \glspl{uav} to improve situational awareness. 
 

In addition, a large number of disaster relief scenarios (e.g. wildfires, earthquakes) or public protection situations (e.g. demonstrations, health hazards) can be served more comprehensively with \glspl{nnpn}. Especially in protracted and large-scale disaster operations that extend over days and weeks, the coordination of operations between different agencies and emergency services can be handled effectively by \glspl{nnpn}.
Figure~\ref{fig:ppdr} illustrates a \gls{ppdr} scenario in which a previous mobile coverage area has been destroyed due to a catastrophic event. Commissioning an \gls{nnpn}, mobile connectivity is re-established, e.g. for emergency services.
With a \textit{Wandering \gls{ran}}, the connectivity is able to move along an arbitrary route. Communication within the cell takes place via the N2/N3 interface, an optional \gls{ntn} backhaul can be used for offloading data.

\begin{figure*}[!ht]
    \centering
    \footnotesize
    \subfloat[Visualization of a \gls{nnpn} in an agriculture scenario.]{
        \includegraphics[width=0.7\textwidth]{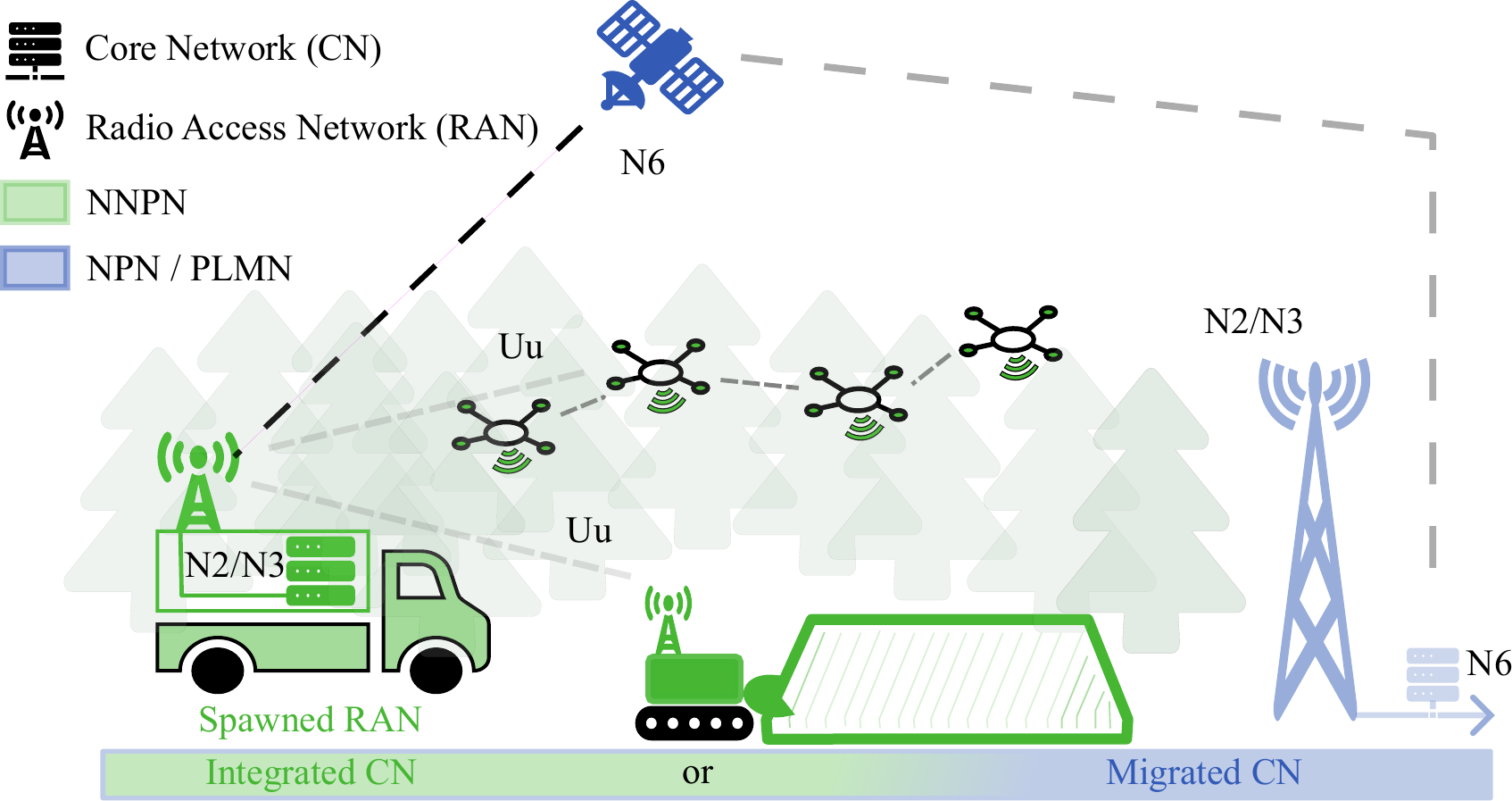}
        \label{fig:agri}}

    \subfloat[Visualization of a \gls{nnpn} in a construction scenario.]{
        \includegraphics[width=0.7\textwidth]{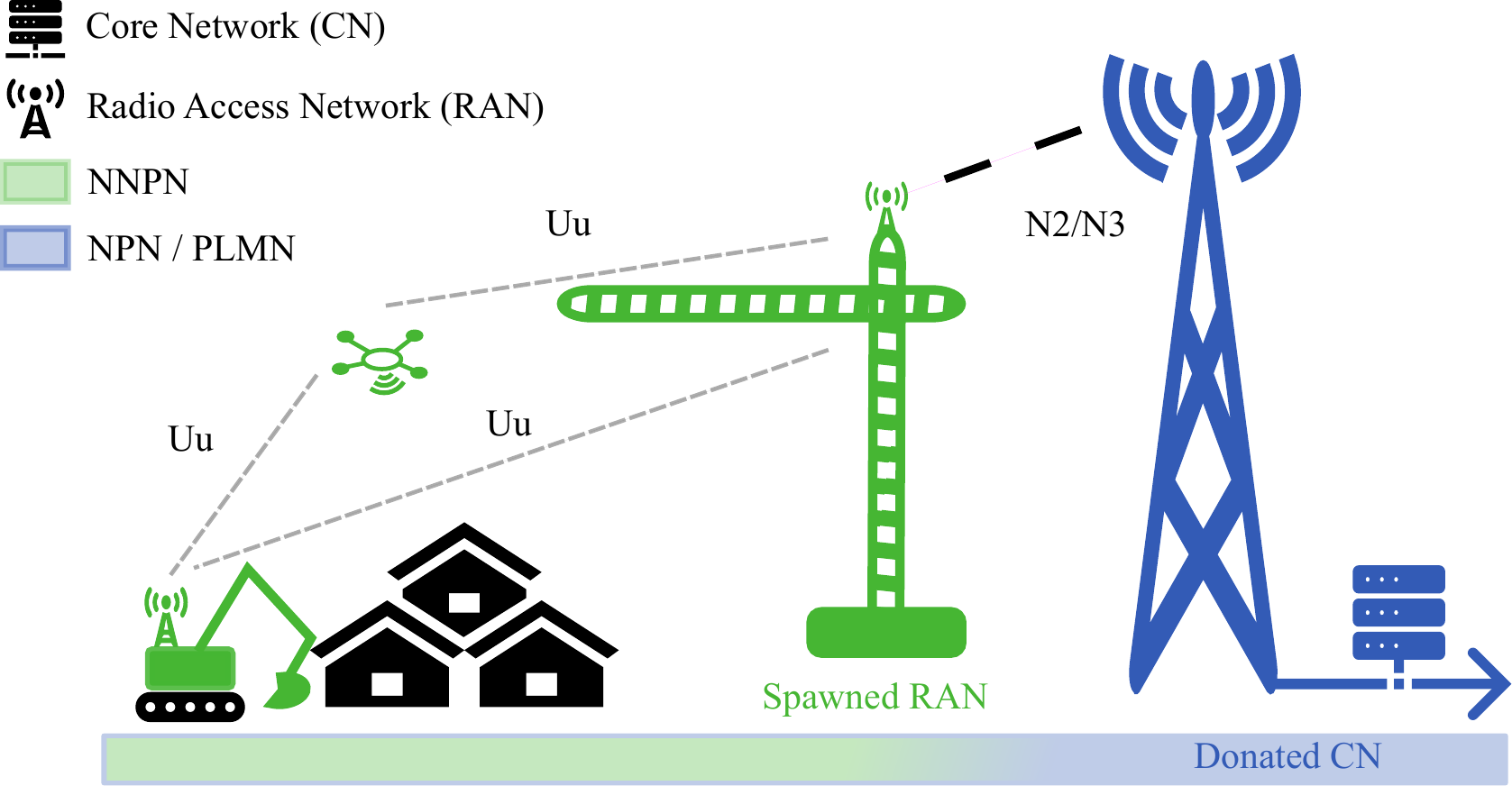}
        \label{fig:const}}
    \caption{Realization scenario of \glspl{nnpn} for agriculture and construction sites}
    \label{fig:agri_const}
\end{figure*}


\subsubsection{Agriculture}
In order to improve the efficiency of farming operations and reduce the impact on the environment, precision agricultural methods will be increasingly used. This involves the ongoing digitalization of farming equipment, which requires a higher level of automatization and autonomous collaboration between machines, which in turn necessitates local networking. For example, a \gls{uav} could autonomously cross a field and transmit videos for further evaluation. 
This evaluation could identify problems such as increased weed growth or nutrient deficiencies. In addition, there are a variety of applications in farming, such as distribution of chemicals or crop monitoring. They are especially beneficial for measuring wide areas and collecting real-time information, which is crucial for high-precision farming.
Figure~\ref{fig:agri} shows a harvesting scenario in which an \gls{nnpn} is used to establish connectivity for the machines in a rural area. The \gls{nnpn} uses a \textit{Spawned \gls{ran}} on a vehicle and an \textit{Integrated \gls{cn}} including \gls{ntn} backhauling via the N6 interface. Depending on the coverage by \glspl{plmn}, a solution via a \textit{Migrated \gls{cn}} is also possible, in which dedicated \glspl{netfunc} are processed via public infrastructure and the N2/N3 interface. 

\subsubsection{Construction sites}
The construction industry involves both a high volume of investment and a considerable demand for payable living space, which must be satisfied in the coming years. Tackling this problem requires, alongside other factors, digitalization and enhanced degree of automation. Robotics is proving to be a key enabler in this transition and its adoption is likely to deliver important benefits for logistics, minimizing construction time, and improving safety and quality assurance. By using \glspl{uav} for land surveying, stakeholders gain unique insight into large parts of the site and its surroundings, enabling a complete evaluation of terrain or topography. 
Precision and speed of collaborative robotics correlates with the performance of the interconnection (latency, jitter). Simultaneously is the physical environment constantly changing, due to the nature of a construction site -- influencing e.g. the coverage -- but also requiring the network to adapt, i.e. to roam along the progress of construction vertically or horizontally. This can be supported via either a spawned or a steered \gls{ran} 
(Figure~\ref{fig:const}), which results in different licensing requirements. In urban areas in particular, it is expected that the construction site is close to public networks. One advantage here is that all \glspl{netfunc} can be processed via a donated \gls{cn}, whereby the N2/N3 interface can also be used for data exchange in this example.

\begin{figure*}[!ht]
    \centering
    \footnotesize
    \includegraphics[width=0.7\textwidth]{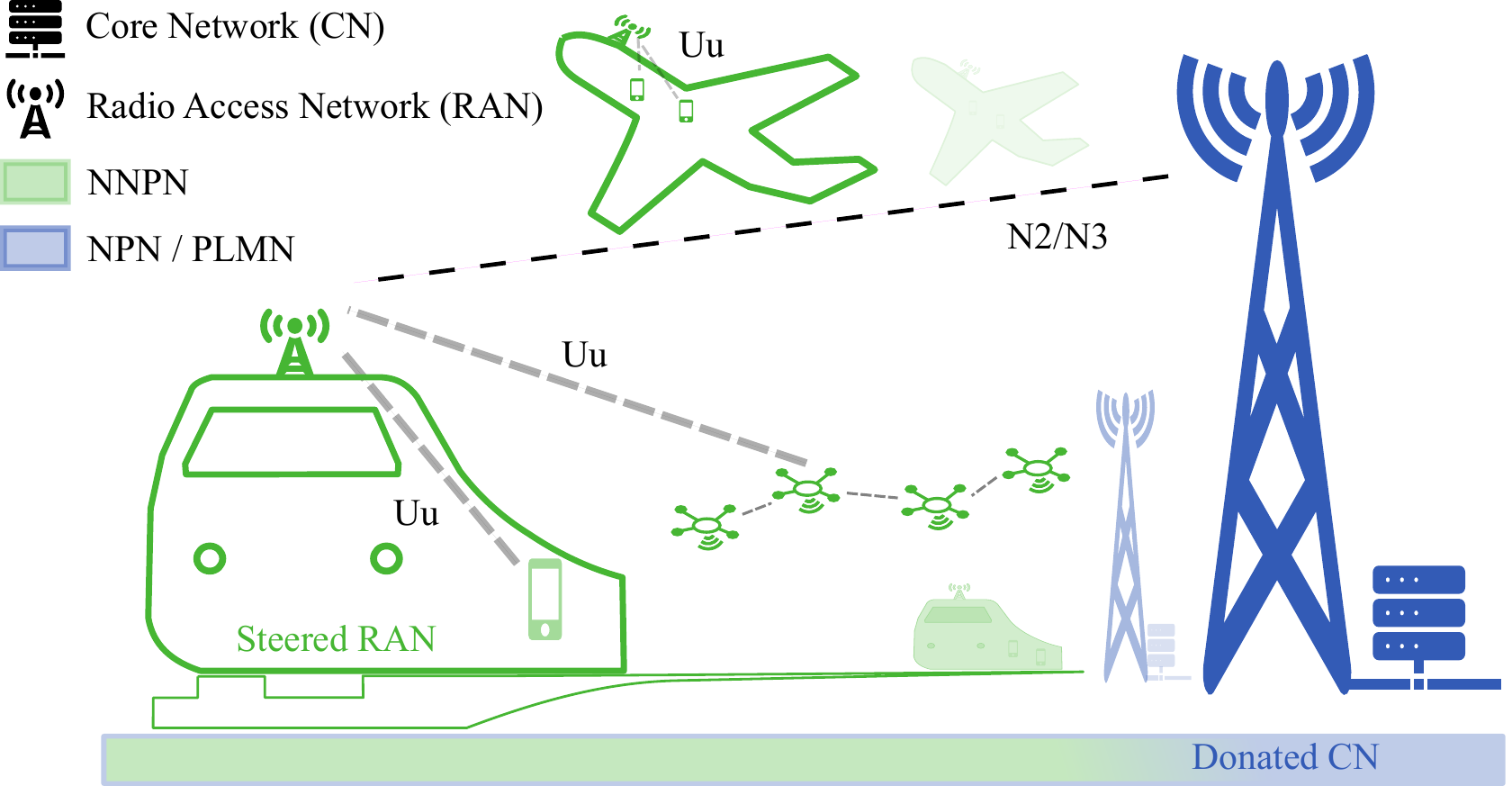}
    \caption{Realization scenario of \glspl{nnpn} for public and private transport}
    \label{fig:transport}
\end{figure*}
\subsubsection{Public and private transport}
In the complex transportation environment, which includes both personal and cargo transport, the networking of every element within these systems is becoming increasingly important. Historically, dedicated solutions customized for different applications, as rail and air transport or logistics, present a challenge as they limit the fluidity of exchange of data between systems. An additional level of difficulty in this area results from the necessity for systems to conform to the different regulatory environments in different countries. For \glspl{nnpn} to work across national borders, global standardization is required. Above all \gls{5g} \gls{npn} have already been successfully tested and introduced in segments of the transport sector. For example, airplanes receive real-time data and weather 
forecast on their systems as they taxi onto the airport runway, which was previously only possible via a cable connection at the boarding gate.
However, the current fixed scenarios in \gls{npn} represent only a small part of the potential use cases and their application area can be significantly extended by using \textit{nomadic} networks. The application of \glspl{uav} for rail safety and railroad inspections is a highly relevant research topic.
In the maritime sector, \glspl{nnpn} can be installed on ships to enable inter-ship communication. The independent \glspl{nnpn} within a fleet connect the ships via an inherently variable topology, enabling both intra- \& inter-fleet communication \cite{TR22.848}.

Fig.~\ref{fig:transport} illustrates a scenario in aviation and rail transport. Here, \glspl{nnpn} can be realized as a steered \glspl{ran} that moves along a defined route. Depending on the availability of \glspl{plmn}, all or parts of the \glspl{netfunc} can be processed via a migrated or donated \gls{cn}.

\subsection{Backhauling}
\label{subsec:backhauling}

Installing a terrestrial and stable backhaul link might be unavailable or impractical in dynamic scenarios or rural areas. An \gls{nnpn} can replace this functionality in an environment where static backhauling cannot adapt to shifting topologies.

In rural areas, where terrestrial infrastructure is expensive to deploy and maintain, \glspl{ntn} can provide high-speed broadband backhauling to existing base stations by acting as intermediate links between the \gls{cn} and smaller sub-networks at the edge, one example being \gls{nnpn}. Satellites and \glspl{hap} can be used to provide an always-on backhaul link \cite{3gpp-ts23.700-27}. For example, a \gls{leo} satellite link can offer individual data rates in the range of 100–200 Mbps -- comparable to current commercial systems like Starlink, which achieve user speeds from approximately 50 to 500 Mbps. In contrast, a \gls{hap}, such as a stratospheric balloon, can provide an aggregate capacity of 1–5 Gbps for a localized area, ensuring robust broadband backhauling. for rural and remote sub-networks.

Using \glspl{isl} for backhauling and dynamic routing of traffic through unaffected terrestrial nodes, these systems enable real-time situational awareness, emergency coordination and remote command and control. For example, satellite-based backhauling can restore connectivity by dynamically allocating bandwidth to the affected region while prioritizing critical traffic, such as video feeds for search and rescue operations. In addition, real-time telemetry from \glspl{ntn} improves decision-making by complementing terrestrial sensing capabilities.

Integrating \gls{ntn} into nomadic networks introduces a number of significant challenges. One major issue is the propagation delay associated with high frequency \gls{ntn} links, such as those operating in the Ka or mmWave bands. These links are particularly sensitive to atmospheric conditions, including rain fade and cloud attenuation, and can suffer from signal degradation over long distances, exacerbated by the dynamic weather conditions found in \gls{ntn} environments. Another critical challenge is the inherently high latency associated with satellite backhauling. For example, \gls{geo} satellite links can introduce one-way propagation delays of 250~ms, making them unsuitable for \gls{urllc} with a round-trip time of ~ 500~ms.

Maintaining proper antenna alignment and managing mobility is also a challenge, as the directional antennas used in \gls{ntn} links require precise alignment. This task becomes even more challenging when the network is nomadic and constantly in motion. As a result, robust handover mechanisms are essential to ensure seamless transitions.

Interference management is another major concern, especially when in-band solutions are deployed. In scenarios where the same frequency bands are used for both access and backhauling links, there is potential for interference that can degrade overall network performance. Efficient resource allocation and advanced interference mitigation techniques are required to minimize this issue. In addition, spectrum limitations and regulatory constraints pose further challenges. The spectrum available for \gls{ntn} backhauling is often subject to strict regulatory requirements and may offer limited bandwidth.

The dynamic topology of networks, driven by the constant movement of satellites and \glspl{hap}, introduces frequent handovers and requires constant reconfiguration of routing paths. This dynamic nature increases control signaling overhead and can lead to temporary service disruption if not managed effectively. Finally, the challenge of balancing backhauling capacity with the need for efficient resource allocation is critical. The link must support both aggregated user data and real-time control signaling, requiring adaptive techniques such as carrier aggregation, load balancing and intelligent resource allocation to maximize throughput while minimizing interference and delay.

\section{Evaluation metrics for \glsfmtshort{nnpn}}
\label{sec:metric}
Evaluating the effectiveness of \glspl{nnpn} requires a structured approach. \glspl{kpi} provide a basis for categorizing \gls{nnpn} use cases by considering factors such as network availability or deployment flexibility. The analysis enables a classification of \glspl{nnpn} and supports their integration into \gls{6g} networks while addressing standardization gaps.

\subsection{Relevant \glsfmtshortpl{kpi}}


To evaluate the effectiveness and customizability of \glspl{nnpn} over the widespread landscape of usage scenarios as well as a metric for categorization,
we defined in Table~\ref{tab:overview_uc} a set of \glspl{kpi}  and collated the mentioned use cases against them:

       \begin{table*}[!b]
            \centering
            \caption{Overview of specific \glspl{kpi} for \gls{nnpn} demands \cite{lindenschmitt2024nomadic}}
            \label{tab:overview_uc}
            \begin{tabular}{|c|c|c|c|c|c|}
            \hline
                           &                &                   &               & \textbf{Construction}  & \textbf{Public/private} \\
                           & \textbf{\glstext{pmse}} & \textbf{\glstext{ppdr}}    & \textbf{Agriculture}   & \textbf{sites}         & \textbf{transport}\\
            \hline
             \textbf{Duration}                                    & hours & hours to days & hours & days to months & minutes to hours\\
            \hline
             \textbf{Schedulability}                                   & possible & no & yes & yes & probably\\
            \hline
             \textbf{Motion predictability}                          & random/planned & random & planned & planned  & planned\\
            \hline
             \textbf{Maximum coverage}                         & micro/pico-cell & micro-cell & micro-cell & pico-cell & pico/femto-cell \\
            \hline
             \textbf{Cross-border interaction}                                & no & possible & no & no & possible\\
            \hline
            \textbf{ Number of active operator }                             & multiple & multiple & single & single & multiple\\
            \hline
             \textbf{Uplink to public network}                    & necessary & necessary & optional & optional & optional\\
            \hline
             \textbf{Safety-relevant communication}                            & possible & yes & no & no & no\\
            \hline
             \textbf{\gls{qos} aspect}                          & \glstext{embb} & \glstext{embb} \& & \glstext{urllc} \& & \glstext{urllc} \& & \glstext{mmtc} \\
                                                        &      & \glstext{urllc}   & \glstext{mmtc}     & \glstext{mmtc}     &      \\
            \hline
             \textbf{Key aspect according to \cite{ITU-R_M.2160-0}}& immersive    & hyper reliable \& & ubiquitous    & ubiquitous    & massive\\
                                                         & communication             & low-latency       & connectivity  & connectivity  & communication\\
                                                         \hline
            \end{tabular}
      \end{table*}
{
\renewcommand{\DenKrDescriptionlabelFormat}[1]{#1}%
\begin{description}
    \item[Duration] 
    defines the required uptime or availability of the \gls{nnpn}, which may range from a few hours to several days, depending on the specific application needs.
    \item[Schedulability] 
    determines whether the network setup can be planned in advance or must be deployed dynamically in response to unforeseen events.
    \item[Motion predictability] 
    describes the movement behavior of an \gls{nnpn}, differentiating between stationary operation, predetermined movement paths, or random spatial mobility.
    \item[Maximum coverage] 
    categorizes network reach into different cell sizes: micro-cells (under 2 km), pico-cells (under 200 m), and femto-cells (under 20 m).
    \item[Cross-border interaction] 
    indicates if the network is confined to a single country or is capable of operating across national borders.
    \item[Number of active operator] 
    specifies whether multiple network operators are active within the \gls{nnpn}'s coverage area, necessitating shared radio resources.
    \item[Uplink to public network] 
    determines if an uplink connection to external networks, such as a cloud server or control center, is required.
    \item[Safety-relevant communication] 
    identifies whether the network facilitates the transmission of safety-critical data, particularly for public protection and emergency response.
    \item[\gls{qos} aspect] 
    highlights key performance parameters such as \gls{embb}, \gls{urllc} or \gls{mmtc}.
    \item[Key aspect according to \cite{ITU-R_M.2160-0}] 
   defines the main aspects consistent with ITU-R M.2160-0, thereby expanding the existing \gls {qos} assessment, with new categories either imposing stricter demands or representing new combinations of existing \gls{qos} requirements (e.g., ubiquitous connectivity corresponds to \gls{mmtc} and \gls{urllc}).
\end{description}

\subsection{Application clusters for \glsfmtshort{nnpn}}
As an outcome of Section \ref{sec:scenarios} and Table \ref{tab:overview_uc} clusters are defined. 

    \begin{itemize}
        \item \gls{uus-nnpn}: 
        Encompasses \glspl{nnpn} that must rapidly adapt to unexpected situations, often requiring coordination among multiple operators within the same coverage area. Typical use cases include \gls{ppdr} and \gls{pmse}.
        \item \gls{ss-nnpn}: 
        refers to \glspl{nnpn} that follow predictable mobility patterns and are managed by a single operator. Common applications include smart agriculture and construction site automation.
        \item \gls{pc-nnpn}: 
        Covers \glspl{nnpn} designed for seamless operation across national boundaries, following predefined movement routes. These networks are primarily used in transportation-related scenarios.
    \end{itemize}

The defined clusters refer to the main variations in the requirements, which means that not all \glspl{kpi} must be covered. In addition, the \gls{qos} and key requirements are very different in their specifications - which on the one hand shows the diverse usability of \glspl{nnpn} and on the other hand means that no clear demarcation is possible. Rather, the specified clusters should be able to cover the requirements defined in \cite{ITU-R_M.2160-0} in specific, as these are seen as an important basis for the further standardization process of \gls{6g}. It is therefore particularly important to point out these gaps in the \gls{5g} standard, which must be closed in order to be able to realize \glspl{nnpn}. 
Even though the availability of an uplink is not necessary from a use case point of view, it can be an asset for individual applications, which is why every cluster should have this capability. Moreover, duration and range were not used for distinguishing the clusters, as these factors are not as relevant for technical considerations as they are for specific realization. As a consequence, the applications are categorized based on their needs and preconditions exist for applications that are not yet addressed.

\section{Realization challenges for \glsfmtshortpl{nnpn}}
\label{sec:challenges}
Integrating \glspl{nnpn} into existing communication systems depends on adaptable network structures, efficient mobility management, and secure connectivity. Additionally, licensing regulations vary by region, having a  significant impact on the realization of \glspl{nnpn}. Understanding these factors is essential for optimizing \glspl{nnpn} performance and enabling their role in \gls{6g}.

\subsection{Architectural Requirements}
\label{subsec:arch_req}
To integrate \glspl{nnpn} effectively, a complete understanding of their requirements is essential. By identifying and organizing these requirements, the complexity of the design process is reduced.

The first challenge addresses the dynamic movement of infrastructure components and the essential capabilities required for mobility. One key aspect is the network architecture, particularly the separation between \gls{cn} and \gls{ran}. In some cases, only specific network elements are mobile, which lowers the energy consumed by moving components and reduces complexity, but in turn increases demands on wireless connectivity. In situations where a whole \gls{ran} is mobile, a reliable wireless backhauling link is necessary through a terrestrial link or an \gls{ntn} link to maintain connectivity with \gls{cn}. These options improve coverage, particularly in remote areas. In addition, computational resources for network services must be allocated dynamically. If an existing infrastructure hosts \gls{cn} functions, coordination between service providers is required to ensure seamless operation.

Another challenge focuses on the operational management of the mobile network's components. Compatibility with legacy systems is essential for an economic and progressive evolution of technologies and commissioned networks. 
Establishing the requirements for \gls{naas} solutions is critical to enable virtualization of \glspl{netfunc} across fixed and mobile infrastructures.  

Collaboration among stakeholders is another key challenge. Service providers, infrastructure managers, and end users contribute to network operations. Effective coordination ensures access control, identity management, and resource allocation. In interference-prone environments, operators must communicate to allocate spectrum efficiently. Furthermore, a nomadic \gls{ran} can enhance network coverage by connecting to an existing \glspl{cn}. 

Security is another important issue. The mobility of \glspl{nnpn} introduces new risks, including unauthorized access and increased attack surfaces. Existing security frameworks may not fully address these threats. Strong authentication, authorization, and data protection mechanisms are necessary. In addition, dynamic trust establishment and secure handover procedures must be implemented to maintain network integrity, resulting in a \gls{taas} solution.

\subsection{Regulatory aspects}
\label{subsec:regulatory}

The current regulatory framework established for \gls{5g} networks does not does not yet address the unique characteristics of \glspl{nnpn}. \gls{5g} introduced \glspl{npn} that operate as independent local private networks. As such, they can be considered a specific form of \glspl{nnpn}. By examining how spectrum for \glspl{npn} is regulated, we can start to understand the challenges \glspl{nnpn} in this area. Spectrum regulation here means organizing, controlling, and monitoring the utilization, allocation, and assignment of frequency bands
.


The allocation of spectrum for local networks varies across countries and influences the future deployment options of \glspl{nnpn}. Several countries have designated different frequency bands for local network use, without an harmonization \cite{matinmikko2023spectrum}. Some frequency bands, such as n258 and n40, are used more frequently, resulting in uniform allocations in neighboring regions, for example in the 26 GHz to 28 GHz range, which is used in many European countries.

The work in \cite{matinmikko2023spectrum} conducts a qualitative analysis for five countries (e.g. USA, Japan, Finland, UK, and Germany) and shows us the diversity of regulations regarding local spectrum assignment (in this case, for the 3.5 GHz band). Finland awarded all the spectrum to existing \glspl{mno} through auctions. Any local network would have to be established by an \gls{mno} or lease the spectrum from them. Germany released the 3.7 GHz band for local networks, and awards spectrum through administrative allocation (i.e. applications can be made to the government for local licenses). Japan has a similar situation, with licenses for local private \gls{5g} indoor networks being issued by the government for a fee. \glspl{mno} are not entitled to request local \gls{5g} permits. The UK also limits the coverage of local networks, licensing only low and medium power base stations under a Shared Access process, on a first-come first-served basis. Finally, the USA has a three-tiered system (\gls{cbrs}), where the lowest tier (referred as General Authorized Access) permits open and flexible access to the band. Devices operating on those frequencies have to avoid interfering with higher-tiered incumbents.

The primary challenge facing \glspl{nnpn} in regard to spectrum is the lack of a quick and straightforward way for the acquisition of exclusive frequency bands. Operators might be located in a region where spectrum is available, but acquiring it can only be done through negotiations with the government or leasing agreements with \glspl{mno}. These spectrum leasing processes are currently conducted manually (meetings, emails, phone calls) \cite{thakur2023streamlining}. The final decision can potentially take weeks or even months. 

Another main challenge appears with cross-border \glspl{nnpn}. Given the national and local nature of spectrum allocation, two neighboring countries can have vastly different regulatory situations. Acquiring the necessary spectrum might entail completely different licensing processes. The network infrastructure needs to have the flexibility to reconfigure themselves with respect to the frequency used to fit the allowed spectrum. This poses a challenge to the hardware used for network deployments. In the event that the available frequency bands have interference and propagation properties that are not suitable for the network services provided, then the \glspl{nnpn} might not be able to be instantiated. Addressing technical and regulatory challenges alike is essential for ensuring the reliable and secure operation of \glspl{nnpn} within static communication networks.

\section{Conclusion}
\label{sec:concl}
This paper analyzed the potential of \glspl{nnpn} in future \gls{6g} communication systems. By extending the concept of \glspl{npn}, \glspl{nnpn} enable mobile and adaptive connectivity for applications such as emergency response, agriculture, and transportation. Through a structured examination of use cases, new \glspl{kpi} like duration, schedulability or motion predictability, metrics, and network classification, this work established a framework for evaluating \gls{nnpn} feasibility especially towards \gls{6g} standardization and to evaluate the efficiency and adoptability of \glspl{nnpn} in various deployment scenarios and emphasize the need for spatially agile \glspl{npn}. As a result, this paper proposes application clusters that cover different needs, ensuring to cover key aspect while enabling variances in \glspl{kpi} and specific application features. 

However, several challenges remain, particularly in spectrum management, dynamic trust management and licensing aspects. The dynamic nature of \glspl{nnpn} requires efficient resource allocation, seamless handovers, and strong authentication mechanisms to maintain reliable performance. Additionally, the integration of \glspl{nnpn} into existing network infrastructures must account for interoperability with legacy systems while ensuring minimal service disruptions. Regulatory concerns, including spectrum availability and cross-border operation, further complicate large-scale deployment and require coordinated policy efforts. In addition, future work should further investigate the requirements of the defined application clusters in relation to the architecture of an \gls{nnpn} and more generally the implications for the standardization of \gls{6g}. 

%
%
\section*{Acknowledgment}%

The authors acknowledge the financial support by the German \textit{Federal Ministry for Education and Research (BMBF)} within the projects »Open6GHub« \{16KISK003K \& 16KISK004\} and »6GTakeOff« \{16KISK067\}.
%

\printbibliography[env=bibliographyDenKrEnv]%

\end{document}